\documentclass[12pt]{iopart}

\usepackage{graphicx}
\usepackage{iopams}
\usepackage{setstack}

\begin{document}

\newcommand{\units}[1]{ \, \mathrm{ #1 } }

\def\Up{U_\mathrm{p}}
\def\rs{r_\mathrm{s}}
\def\ts{t_\mathrm{s}}
\def\omegaM{\omega_\mathrm{Mie}}

\def\beq{\begin{eqnarray}}
\def\eeq{\end{eqnarray}}

\title{Collective-field-corrected strong field approximation for laser-irradiated metal clusters}

\author{Th.\ Keil and D.\ Bauer}
\address{Universit\"at Rostock, Institut f\"ur Physik, 18057 Rostock, Germany}

\begin{abstract}
The strong field approximation (SFA) formulated in terms of so-called ``quantum orbits'' led to much insight into intense-laser driven
 ionization dynamics. In plain SFA, the emitted electron is treated as a free electron in the laser field alone. However, with improving experimental techniques and more advanced numerical simulations it becomes more and more obvious that the plain SFA misses interesting effects even on a qualitative level. Examples are holographic side lobes, the low-energy structure, radial patterns in photoelectron spectra at low kinetic energies, and strongly rotated angular distributions. For this reason increasing effort has been recently devoted to Coulomb corrections of the SFA. In the current paper, we follow a similar line but consider ionization of metal clusters. It is known that photoelectrons from clusters can be much more energetic than those emitted from atoms or small molecules, especially if the Mie resonance of the expanding cluster is evoked. We develop a SFA that takes the collective field inside the cluster via the simple rigid-sphere model into account. Our approach is based on field-corrected quantum orbits so that the acceleration process (or any other spectral feature of interest) can be investigated in detail. 

\end{abstract}

\submitto{\jpb}

\maketitle

\section{Introduction}
The strong field approximation (SFA) \cite{ref_KFR_amplitude} is {\em the} underlying theory describing the interaction of intense laser light with atoms \cite{ref_JPB_review} or molecules \cite{milo_molecules}. It allows to calculate photoelectron \cite{ref_JPB_review} or high-harmonic spectra \cite{lew_hohg} and provides a deep understanding of the cut-offs and the interferences observed in these spectra. However, it has been noted that several spectral features are not reproduced by the plain SFA, examples being radial structures at low photoelectron energies \cite{ref_Rudenko_radial_structure,ref_Arbo_radial,ref_PRL_Gopal} holographic side-lobes \cite{ref_Science_Huismans}, or the low-energy structure \cite{ref_Nature_Phys_LES,ref_PRL_Quan}. The reason for this failure of the SFA is the neglect of the Coulomb force on the outgoing (or returning) photoelectron. Attempts to include Coulomb effects have a long history \cite{ref_Popov_Coulomb_correction,Popov_review} and continue to date \cite{ref_CVA_Faisal,ref_CVA_Arbo,ref_EVA,ref_JMO_Popruzhenko,ref_Coulomb_asymmetry_circular_CCSFA,ref_PRL_TCSFA,ref_DDCV_Arbo_Ciappina,torlinaI}, mainly because more and more advanced detector technology  and {\em ab initio} simulations reveal more and more features {\em not} covered by the plain SFA. In general, the more ``differential'' the observable is, the trickier is the proper inclusion of Coulomb effects. While the simplest task is to correct total ionization rates \cite{ref_Popov_Coulomb_correction,Popov_review}, much more challenging, for instance, is to get the interference pattern in photoelectron momentum spectra right \cite{tmy_latest}.       

It is well-known that photoelectrons generated by strong-field ionization of atoms have a cut-off energy of $2\Up$ if they move directly to the detector (or $10\Up$ if they rescatter once from their parent ion) (see, e.g., \cite{ref_JPB_review,mubaubook} for reviews). These are the cut-offs predicted by the plain SFA (or the SFA extended for one rescattering event, respectively), which are hardly affected by the Coulomb correction. On the other hand, electrons emitted from laser-irradiated clusters may have much higher kinetic energies, especially at resonance \cite{sparc}. In this work, we apply the methodology developed for the Coulomb-corrected SFA based on quantum trajectories \cite{ref_JMO_Popruzhenko,ref_Coulomb_asymmetry_circular_CCSFA,ref_PRL_TCSFA,tmy_latest,tmy_puils} to a SFA that is corrected for the collective electric field in the cluster. In this way we are able to show that it is this collective field, which arises because of the coherent oscillation of the electron cloud with respect to the ionic background, that generates multi-$\Up$ electrons. This finding confirms the SPARC effect (i.e., ``surface-plasmon-assisted rescattering in clusters'') revealed earlier via classical molecular dynamics simulations \cite{sparc}.  The aim of the current paper is to introduce a quantum, SFA-based method for metal clusters that treats the ionization step self-consistently and allows for interference effects. Similar approaches may then be also applied to other situations where collective fields play a role, for instance electron emission from metal nanotips \cite{nanotipNature,quantumorbitfieldenhancement}.

The paper is organized as follows. In section \ref{sec:theory}, the basic ingredients, i.e., the SFA, field-corrected quantum orbits, the rigid sphere model for clusters, and the actual numerical implementation  are introduced or reviewed.  In section \ref{sec:results} 
a typical photoelectron spectrum obtained for a Na$_{20}$-cluster close to resonance is presented, and the origin of the fast electrons is revealed by identifying the relevant, collective-field-corrected quantum orbits. Finally, we summarize in section \ref{sec:concl}. The equation of motion for the electron sphere in the rigid sphere model is derived in \ref{app:restoringforce}. 

Atomic units  are used (in which, numerically, $\hbar =m_e=|e|=4\pi\varepsilon_0=1$) unless otherwise noted.

\section{Theory}
\label{sec:theory}
\subsection{Rigid sphere model (RSM)}
\label{sec:rigidspheremodel}
\begin{figure}
  \includegraphics{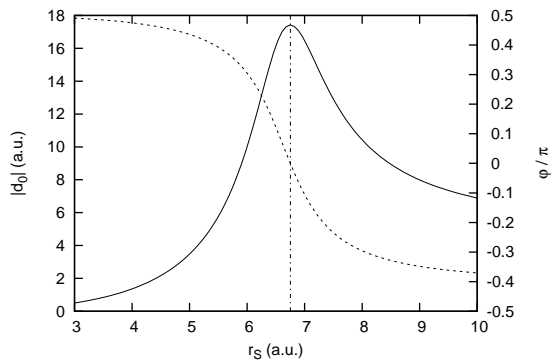}
  \caption{Excursion amplitude  $|d_0|$ (solid line, left axis) and $\varphi$ (dotted line, right axis) according \eref{eq:oscillationparameters} as functions of the Wigner-Seitz radius $\rs=\omegaM^{-2/3}$ for $E_0=0.01688$, $\omega=0.057$, $\gamma = 0.017$  (see  \sref{sec:results}). The dash-dotted vertical line denotes the $\rs$ where resonance occurs. }
  \label{fig:oscillationparameters}
\end{figure}

The collective field by which the quantum orbits of the SFA will be corrected is approximated using the  RSM. 
In the RSM both electrons and ions are modeled by homogeneously charged spheres. We assume that the radii and the absolute values of the charges of the electron and the ion sphere are equal. Further, we assume one valence electron per atom (as in sodium Na$_N$ clusters) so that the cluster radius reads $R=N^{1/3}\rs$, with $\rs$ the Wigner-Seitz radius ($\rs\simeq 4$ for bulk sodium). Both ion and electron number densities are $n_0 = 3/(4\pi \rs^3)$. 
Driving this system with an external (laser) field in dipole approximation $\bi{E}(t)$  leads to the following equation of motion  (cf.\  \ref{app:restoringforce} for a derivation)
\begin{eqnarray}
  \label{eq:displacementeomanharmonic}
\ddot \bi{d} = -\omegaM^2  \left( \bi{d} - \frac{9}{16R}\bi{d}|\bi{d}| + \frac{1}{32R^3} \bi{d}|\bi{d}|^3 \right) - \bi{E} - \gamma \dot \bi{d} . 
\end{eqnarray}
Here, $\bi{d}(t)$ is the displacement of the center of the electron sphere with respect to the center of the ion sphere (located in the origin), and
\beq 
\omegaM = \sqrt{\frac{4\pi n_0}{3}}=\rs^{-3/2}
\eeq 
is the Mie frequency. Equation \eref{eq:displacementeomanharmonic} is the equation of motion for a driven, damped, anharmonic oscillator. A damping $\gamma$ is introduced to prevent the singularity in the excursion amplitude at resonance (see below). If the laser polarization is linear, e.g., $\bi{E} \parallel \bi{e}_z$, the excursion $\bi{d}$ will be along $\bi{e}_z$ as well.  For small excursions of the electron sphere $d_z=\bi{e}_z\cdot \bi{d}$, $|d_z| \ll R$  the harmonic oscillator term $\sim \bi{d}$ will dominate, and the anharmonicities $\sim \bi{d}|\bi{d}|$  and $\sim \bi{d}|\bi{d}|^3$ can be neglected. Then, for an  electric field of the form 
\beq \bi E(t)=\bi{e}_z E_0\cos \omega t\eeq
the excursion and phase are 
\begin{eqnarray}
  \label{eq:oscillationparameters}
  d_z(t) = d_0 \sin(\omega t + \varphi), \quad d_0=\frac{-E_0}{\sqrt{( \omega_{\mathrm{Mie}}^2 - \omega^2 )^2 + \gamma^2 \omega^2 }},\nonumber\\
  \varphi = \arctan \left( \frac{ \omega_{\mathrm{Mie}}^2 - \omega^2 }{ \gamma \omega } \right).
\end{eqnarray}
 Excursion amplitude  $d_0$ and phase $\varphi$ as functions of the Wigner-Seitz radius $\rs=\omegaM^{-2/3}$ are shown in \fref{fig:oscillationparameters}. As expected, for $\omegaM^2\ll\omega^2$ (i.e., $1\ll \omega^2\rs^3$) and $\gamma \ll \omega$ the free-electron limit $\varphi=-\pi/2$, $d_z(t) = E(t)/\omega^2$ is obtained. In the opposite limit   $\omegaM^2\gg\omega^2$,  $\gamma \ll \omegaM$ one has  $\varphi=\pi/2$, $d_z(t) = -E(t)/\omega^2$. At resonance, $\varphi=0$, which occurs for the laser parameters chosen at $\rs=6.75$.
For finite laser pulses with an envelope $E_0(t)$ the general analytical solution including transient effects is more involved than \eref{eq:oscillationparameters}. For the purpose of this paper it is sufficient to apply \eref{eq:oscillationparameters} adiabatically, i.e.,   the constant laser amplitude $E_0$ is replaced by $E_0(t)$. 

\begin{figure}
  \includegraphics{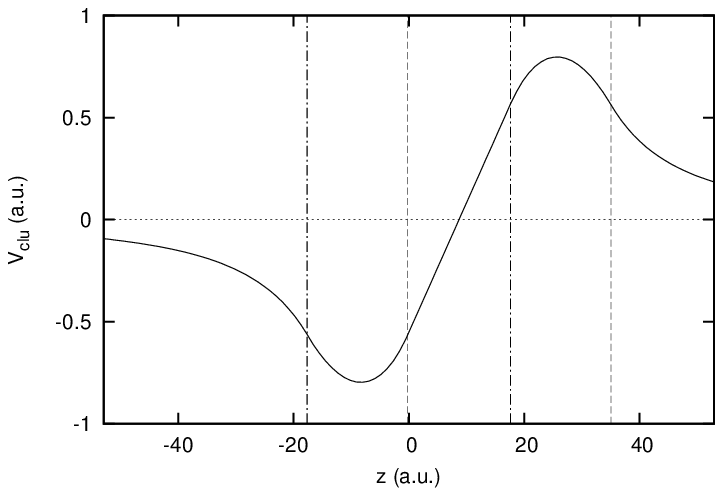}
  \caption{The cluster potential $V_{\mathrm{clu}}(z)$ (solid line) from \eref{eq:singleelectronclusterpotential} at maximum elongation of the electron cloud (at resonance) using the parameters given in the caption of \fref{fig:oscillationparameters}.  The dash-dotted lines denote the boundaries of the ion distribution, the dashed lines show the boundaries of the electron cloud. Note the steep part (i.e., strong force) in the region where both spheres overlap. }
  \label{fig:oscillationpotential}
\end{figure}

Ion and electron sphere, when centered in the origin, give rise to the same spherical electrostatic potential but with opposite sign,
\begin{eqnarray}
  \label{eq:spherepotential}
  V^{\pm}_{\mathrm{sphere}}(r) = \pm \frac{R^3}{\rs^3} \cases{\frac{r^2}{2R^3}-\frac{3}{2R}  & for\ \ $ r < R$ \\ - \frac{1}{r}  & for\ \ $ r\geq R$}\,.
\end{eqnarray}
 This expression, together with the excursion of the electron sphere $\bi{d}(t)$ in \eref{eq:oscillationparameters},  can be used to construct the cluster potential in  which a test electron would move, 
\begin{eqnarray}
  \label{eq:singleelectronclusterpotential}
  V_{\mathrm{clu}}(\bi{r},t)=V^{+}_{\mathrm{sphere}}(|\bi{r}|) + V^{-}_{\mathrm{sphere}}(|\bi{r}-\bi{d}(t)|)\,.
\end{eqnarray}
This potential is shown in \fref{fig:oscillationpotential} and used for the cluster correction in section \ref{sec:clustercorrection}.\footnote{One could take into account the effect that there should be (at least) one net missing electron charge in $ V_{\mathrm{clu}}$, namely the one of the emitted electron. However, for the acceleration mechanism explored in the current work this effect is negligible (as long as the net charge of the cluster is not too high).} 

\subsection{Strong field approximation (SFA)}
Within the plain SFA the binding force on the outgoing electron is neglected because the electron is described by the solution of the time-dependent Schr\"odinger equation for a free electron in a (laser) field but without binding potential, a so-called Gordon-Volkov state \cite{ref_GV}
\begin{equation}
  |\Psi_{\bi{p}}^{\mathrm{(V)}}(t)\rangle = \rme^{-\rmi S_{\bi{p}}(t)} |\bi{p}+\bi{A}(t)\rangle . \label{eq:GVstate}
\end{equation}
Here, $\bi{A}(t)$ is the vector potential of the laser field, $\bi{E}(t) = -\partial_t \bi{A}(t)$, $\bi{p}$ is the canonical momentum (which is related to the electron velocity $\bi{v}_0$ by $\bi{v}_0=\bi{p}+\bi{A}$), and 
\begin{equation}
  S_{\bi{p}}(t)=\int^t \frac{1}{2}[\bi{p}+\bi{A}(t')]^2\rmd t'
\end{equation}
is the action.
The  Gordon-Volkov state \eref{eq:GVstate} is given in length gauge (although expressed in terms of the canonical momentum and the vector potential).
The plain SFA transition matrix element in length gauge for the so-called direct electrons (i.e., those that move directly to the detector, without hard rescattering) reads (see, e.g., \cite{ref_JPB_review,mubaubook} for reviews)
\begin{equation}
  M_{\bi{p}}^{\mathrm{(SFA)}} = -\rmi \int_0^{\infty} \langle\bi{p}+\bi{A}(t)| \bi{r} \cdot \bi{E}(t) | \Psi_0 \rangle \rme^{-\rmi S_{\bi{p},I_{p}}(t)} \rmd t \label{eq:plainSFAmatrixelem}
\end{equation}
where 
\begin{equation}
  \label{eq:modifiedactionsfa}
  S_{\bi{p},I_p}(t)=\int^t \frac{1}{2}[\bi{p}+\bi{A}(t')]^2 + I_{p}\rmd t'
\end{equation}
is the action including the field-free evolution of the initial bound state $|\Psi_0(t)\rangle=\rme^{\rmi I_p t }|\Psi_0\rangle$ with ionization potential $I_p$.  

\subsection{Quantum trajectory method}
\label{sec:trajectorymethod}
The plain SFA matrix element \eref{eq:plainSFAmatrixelem} for given $\bi{A}(t)$, $|\Psi_0\rangle$, and $I_p$  can easily be evaluated numerically. However, not much insight into {\em why} certain spectral features in $\left| M_{\bi{p}}^{\mathrm{(SFA)}}\right|^2$ appear is gained in that way. In the case of $I_p/\omega \gg 1$  the method of steepest descent can be applied to the time integral in  \eref{eq:plainSFAmatrixelem} \cite{ref_JPB_review}.  The matrix element is then represented by a sum over all saddle point times $\ts^{(\alpha)}$ 
\begin{equation}
  M_{\bi{p}}^{\mathrm{(SFA)}} = \sum_{\alpha} f_{\Psi_0}(\bi{p},I_p,\ts^{(\alpha)})\,  \rme^{-\rmi S_{\bi{p},I_{p}}(\ts^{(\alpha)})} . 
\end{equation}
Here, $f_{\Psi_0}(\bi{p},I_p,\ts^{(\alpha)})$ is a pre-exponential factor that depends on the initial state $|\Psi_0\rangle$ and is evaluated at the respective saddle point time $\ts^{(\alpha)}$. The overall qualitative structure of the photoelectron spectra is not affected by the pre-exponential factor $f_{\Psi_0}(\bi{p},I_p,\ts^{(\alpha)})$. In particular, cut-offs are determined by the exponential $\rme^{-\rmi S_{\bi{p},I_{p}}(\ts^{(\alpha)})}$, not by $f_{\Psi_0}(\bi{p},I_p,\ts^{(\alpha)})$.\footnote{In our actual implementation we tested, besides $f \equiv 1$, several $|\Psi_0\rangle$ (hydrogenic, Gaussians of width $\sim R$, short-range $\delta$-potential-like). The results shown in the following are calculated for hydrogenic $|\Psi_0\rangle$ \cite{ref_JPB_review}.}

The  saddle point times $\ts^{(\alpha)}$ for a given momentum $\bi{p}$ are determined by
\begin{equation}
  \label{eq:saddlepointcondition}
  \left. \frac{\partial S_{\bi{p},I_{p}}}{\partial t} \right|_{\ts^{(\alpha)}} = 0 \qquad \Rightarrow \qquad \frac{1}{2} [\bi{p}+\bi{A}(\ts^{(\alpha)})]^2 = -I_p,
\end{equation}
which can be interpreted as energy conservation at the time instant of ionization. Since $\bi{p}$ is real and $I_p$ is positive, the saddle point times $\ts^{(\alpha)}$ are necessarily complex. Every saddle point time represents a so-called quantum trajectory $\bi{r}(t)$ \cite{ref_quantum_orbits,ref_Scinece_Feynmann,ref_PRL_time_slits,ref_paulusbauer} which starts in the complex time plane  at $\ts=\ts^{(\alpha)}$. Typically two saddle point times per cycle contribute to a given $\bi{p}$.\footnote{These are  the so-called ``short'' and ``long'' trajectory.}  The propagation obeys  Newton's equation of motion but in complex space and time. The plain-SFA quantum trajectories are thus obtained by
\begin{equation}
  \bi{r}(t)=\int_{\ts}^t[\bi{p}+\bi{A}(t')]\rmd t' + \bi{r}(\ts)\,,
\end{equation}
with initial conditions for $\bi{r}(\ts)$ still to be chosen. The plain-SFA matrix element $\left| M_{\bi{p}}^{\mathrm{(SFA)}}\right|^2$ does not depend at all on these initial conditions. However, the corrected SFA matrix element to be introduced below does. In the atomic case one may choose  $\mathrm{Re}\,\bi{r}(\ts)=0$ (i.e., the real part of the electron quantum orbit starts at the position of the nucleus). As a consequence,  $\bi{r}(\ts)$ has to be purely imaginary, $\bi{r}(\ts)=\rmi \, \mathrm{Im}\,\bi{r}(\ts)$. The other condition is given by the canonical momentum $\bi{p}$ which fixes the initial velocity
\begin{equation}
  \dot{\bi{r}}(\ts)=\bi{v}_0(\ts)=\bi{p} + \bi{A}(\ts)\,.
\end{equation}
The electron reaches the classically allowed region at $t_r=\mathrm{Re}\,\ts$.  The so-called ``tunnel exit'' $\bi{r}(t_r)$ thus reads
\begin{eqnarray}
  \bi{r}(t_r) = \bi{r}(\mathrm{Re}\,\ts) &= \int_{\ts}^{t_r}[\bi{p}+\bi{A}(t')]\rmd t' +\bi{r}(\ts)\\
  &= -\rmi\,\bi{p}\,\mathrm{Im}\,\ts + \bi{a}(t_r) - \bi{a}(\ts) + \rmi\,\mathrm{Im}\,\bi{r}(\ts)
\end{eqnarray}
with the excursion of a free electron in a laser field
\begin{equation}
  \bi{a}(t) = \int^t \bi{A}(t')\rmd t'\,.
\end{equation}
We choose the initial position $\bi{r}(\ts)=\rmi\,\mathrm{Im}\,\bi{r}(\ts)$ such that the tunnel exit $\bi{r}(t_r)$ is real,
\begin{equation}
  \label{eq:tunnelexitsfa}
  \bi{r}(t_r)=\bi{a}(t_r) - \mathrm{Re}\,\bi{a}(\ts)\,,
\end{equation}
which leads to a real position 
\beq
\bi{r}(t) = \int_{t_r}^t[\bi{p}+\bi{A}(t')]\rmd t' + \bi{r}(t_r) , \qquad t \geq t_r 
\eeq 
for the electron for all real times $t \geq t_r\,$.\footnote{There are also good arguments against such a choice of purely real $\bi{r}(t)$ ``after'' the tunnel exit \cite{torlinaII}, e.g., concerning the analyticity of $\bi{r}(t)$. }

\subsection{Cluster correction}
\label{sec:clustercorrection}
The quantum trajectory approach offers a convenient way to correct the SFA for the effect of external potentials on the outgoing electron.  Without external potential the canonical momentum $\bi{p}$ is conserved in a laser field in dipole approximation. This fact can be used to recast the action \eref{eq:modifiedactionsfa} 
into the form
\begin{equation}
  \label{eq:matrixelementtrajectoriesrecast}
  S_{\bi{p},I_p}(\ts^{(\alpha)})=C(\bi{p}) - \int_{\ts^{(\alpha)}}^{\infty}\left[ \frac{1}{2}\bi{v}_0^2(t) + I_p \right]\rmd t
\end{equation}
with the electron velocity $\bi{v}_0(t)=\bi{p} + \bi{A}(t)$ and the purely momentum-dependent term $C(\bi{p}) = \int_0^\infty \frac{1}{2}[\bi{v}_0^2(t)+I_p]\rmd t$. This term factors out of the coherent summation of contributions for a fixed asymptotic momentum $\bi{p}$ since it does not change with the saddle-point time $\ts^{(\alpha)}$. Therefore it does not contribute to the final ionization probability.

We rewrite the second term in \eref{eq:matrixelementtrajectoriesrecast} using the Gordon-Volkov Hamiltonian of a free electron in a laser field $H_0(t) = \frac{1}{2}\bi{v}_0^2(t) = \frac{1}{2}[\bi{p}+\bi{A}(t)]^2\,$,
\begin{equation}
  S_{\bi{p},I_p}(\ts^{(\alpha)})=C(\bi{p}) - \int_{\ts^{(\alpha)}}^{\infty}\left[ H_0(t) + I_p \right]\rmd t\,.
\end{equation}
The actual correction due to the cluster potential \eref{eq:singleelectronclusterpotential} now reads
\begin{equation}
  H_0(t)\,\rightarrow\,H(t)=\frac{1}{2}\bi{v}_{\mathrm{corr}}^2(t)+V_{\mathrm{clu}}(\bi{r},t)\,.
\end{equation}
Because of this modification the canonical momentum is not conserved anymore.  The corrected velocity becomes 
\beq \bi{v}_{\mathrm{corr}}(t)=\bi{p}_{\mathrm{corr}}(t)+\bi{A}(t)\eeq 
with $\bi{p}_{\mathrm{corr}}(t)$ 
the corrected (now time-dependent) canonical momentum. The asymptotic momentum $\bi{p}_{\mathrm{final}}$ for every $\ts^{(\alpha)}$ needs to be calculated by numerical propagation of the associated trajectory.
The final result for the transition amplitude then reads
\beq
   M_{\bi{p}}^{\mathrm{(cluster)}} = \sum_{\alpha} M_{\bi{p}}(\ts^{(\alpha)})  \label{eq:correctedtransitionamplitude}
\eeq
with
\beq  M_{\bi{p}}(\ts^{(\alpha)}) = f_{\Psi_0}\,\rme^{-\rmi S_{\mathrm{corr}}(\ts^{(\alpha)})},\eeq
 \beq S_{\mathrm{corr}}(\ts^{(\alpha)})=\int_{\ts^{(\alpha)}}^\infty \left[ \frac{1}{2}\bi{v}_{\mathrm{corr}}^2(t) + V_{\mathrm{clu}}(\bi{r},t) + I_p \right] \rmd t .\label{eq:correctedtransitionamplitudeS} \eeq

\subsection{Numerical implementation}
The actual evaluation of the transition amplitude \eref{eq:correctedtransitionamplitude} is done in several steps. First, the saddle-point equation 
\begin{eqnarray}
  \frac{1}{2} (\bi{p}_{\mathrm{initial}}+\bi{A}(\ts^{(\alpha)}))^2 = -I_p\,,
\end{eqnarray}
is solved for randomly or uniformly ``shot'' initial canonical momenta $\bi{p}_{\mathrm{initial}}$ within the momentum range of interest. We use a complex-root-finding algorithm for finding all $\ts^{(\alpha)}$ for a given $\bi{p}_{\mathrm{initial}}$.\footnote{We use the ACM TOMS algorithm 365 \cite{toms365}.} 
 Every solution $\ts^{(\alpha)}$ corresponds to one trajectory for which  the time integral in $S_{\mathrm{corr}}(\ts^{(\alpha)})$  needs to be solved. The latter is splitted according
\begin{eqnarray}
  \label{eq:matrixelementtimeintegral}
  S_{\mathrm{corr}}(\ts^{(\alpha)}) \simeq S_\mathrm{sub-barrier} + S
\end{eqnarray}
where
\begin{eqnarray}
S_\mathrm{sub-barrier}=\int_{\ts^{(\alpha)}}^{t_r^{(\alpha)}} \left[ \frac{1}{2}\bi{v}_0^2(t) + I_p \right] \rmd t
\end{eqnarray}
and
\begin{eqnarray}
S=\int_{t_r^{(\alpha)}}^\infty \left[ \frac{1}{2}\bi{v}_{\mathrm{corr}}^2(t) + V_{\mathrm{clu}}(\bi{r},t) + I_p \right] \rmd t . \label{eq:matrixelementtimeintegralS}
\end{eqnarray}
The cluster correction is neglected during the propagation in complex time, i.e., in $S_\mathrm{sub-barrier}$ (during the so-called ``sub-barrier motion''). For a given vector potential $\bi{A}(t)$ this part can thus be solved analytically. This avoids the solution of  Newton's equations of motion in complex space and time as we correct the electron trajectory only from the tunnel exit to the detector. The integral in $S$ is over real times only but needs to be  computed numerically because of the presence of $V_{\mathrm{clu}}(\bi{r}(t),t)$. The trajectory for a certain $t_r^{(\alpha)}=\mathrm{Re}\,\ts^{(\alpha)}$ is calculated according to Newton's equations of motion in real space and time (dropping the subscript 'corr' for brevity),
\begin{eqnarray}
  \label{eq:newtonsequationscorrected}
  \bi{v}(t) = \dot{\bi{r}}(t) = \bi{p}(t)+\bi{A}(t), \nonumber \\
  \dot{\bi{p}}(t) = -\bnabla V_{\mathrm{clu}}(\bi{r}(t),t), \label{eq:eoms}\\
  \dot{S}(t) = \frac{1}{2}\bi{v}^2(t) + V_{\mathrm{clu}}(\bi{r}(t),t) + I_p,\nonumber
\end{eqnarray}
subject to the initial conditions
\beq \bi{p}(t_r^{(\alpha)}) = \bi{p}_\mathrm{initial}, \qquad  \bi{r}(t_r^{(\alpha)})=\bi{a}(t_r^{(\alpha)}) - \mathrm{Re}\,\bi{a}(\ts^{(\alpha)}), \qquad  S(t_r^{(\alpha)})=0. \eeq
This set of ordinary, coupled differential equations of first order is propagated numerically from $t_r^{(\alpha)}$ to $t = t_{\mathrm{final}}$ for a sufficiently large $t_{\mathrm{final}}$ such that it is ensured that $\bi{p}(t)$ for $t > t_{\mathrm{final}}$ does not change significantly anymore.\footnote{In the case of a pure Coulomb potential the asymptotic momentum $\bi{p}$ can be calculated analytically once the laser is off.} Including $\dot S$ in the set of differential equations \eref{eq:eoms} is an efficient way to calculate $S$ in \eref{eq:matrixelementtimeintegralS}.  From the results the individual transition amplitude $M_{\bi{p}}(\ts^{(\alpha)})$ for every $\ts^{(\alpha)}$ can be calculated. To obtain a full momentum spectrum $\left|M_{\bi{p}}^{\mathrm{(cluster)}}\right|^2$ the steps described above are carried out for many uniformly or randomly shot initial momenta $\bi{p}_\mathrm{initial}$.    The resulting trajectories are binned according  their final momenta $\bi{p}(t_\mathrm{final})$. In order to calculate $M_{\bi{p}}^{\mathrm{(cluster)}}$ the individual transition amplitudes of the trajectories in a final-momentum bin are added up coherently. In that way interference effects are incorporated in $\left|M_{\bi{p}}^{\mathrm{(cluster)}}\right|^2$.\footnote{Interference of quantum trajectories underlies the idea of holographic imaging by photoelectrons \cite{ref_Science_Huismans}. The acceleration mechanism in clusters discussed in this work, however, does not rely on interference.}

\section{Results and discussion}
\label{sec:results}
We consider a sodium cluster consisting of $20$ atoms in a three-cycle,  $\sin^2$-shaped laser pulse with a frequency of $\omega=0.057$ (corresponding to $800$\,nm) and an amplitude of the electric field of $E_0=0.01688$ (corresponding  to a laser peak intensity of $I_0=10^{13}\units{W/cm^2}$). The oscillation of the electron cloud with respect to the ions is described by the RSM introduced in \sref{sec:rigidspheremodel}. The only free parameter in this model is the damping factor $\gamma$. In the real cluster system damping occurs because of electron-ion collisions, Landau damping, and emission of electrons from the cluster as a whole. For the simulation we have chosen $\gamma= 0.017 < \omega/3$ as small as possible but big enough to ensure that even at resonance electron and ion sphere always overlap. Allowing for larger excursions in the RSM---while assuming an unaffected shape of the electron sphere---seems unreasonable.

As mentioned already in \sref{sec:trajectorymethod}, the choice of $|\Psi_0\rangle$ only affects  the pre-exponential factor and thus does not change the qualitative features of the photoelectron spectra, in particular the cut-offs. Instead, the ionization potential $I_p$ enters the exponential. We set it to the experimental value $I_p=0.14$ for the ground state of Na$_{20}$ \cite{EiPi}.  Changes of the ionization potential due to cluster expansion are neglected. Another ingredient in the cluster-SFA is the position of the tunnel exit $\bi{r}(t_r)$ \eref{eq:tunnelexitsfa}, which has to be modified to make sure that an emitted electron appears {\em outside} the cluster, as is the case in tunneling ionization. To that end  the atomic tunnel exit is shifted outwards by three times the cluster radius. Along the laser polarization direction $\bi{e}_z$ we thus have $z(t_r) \to z_\mathrm{cluster}(t_r)= z(t_r) + 3 R \, z(t_r)/|z(t_r)| $. This procedure seems rather {\em ad hoc}. However, the photoelectron spectra are robust with respect to changes in this shift as long as (i) the electron starts outside the cluster, i.e., $|z_\mathrm{cluster}(t_r)| > |d_z(t_r)|+R $, and (ii) the laser is still on when the above mentioned long SFA-orbits pass through the cluster center.

Pump-probe schemes are used in the experiment \cite{sparc} to allow the clusters to expand.  The first pulse excites the cluster, the second one drives the ionization. Between the pulses, the cluster expands, changing its density and thus the Mie frequency. This behavior is mimicked in the simulations by a variation of the Wigner-Seitz radius $\rs$ of the cluster. The pump pulse itself is not simulated.

\subsection{Photoelectron momentum spectra}
\begin{figure}
  \includegraphics{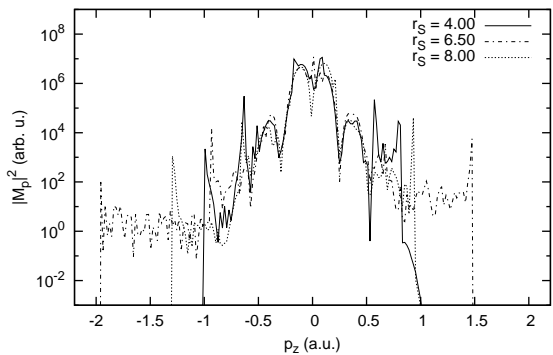}
  \caption{Photoelectron momentum spectra for different values of $\rs$, a  three-cycle,  $\sin^2$-shaped laser pulse with a frequency of $\omega=0.057$  and an amplitude of the electric field of $E_0=0.01688$. For the unexpanded cluster, $\rs=4$ (solid line), the Mie-frequency is greater than the laser frequency, the resonance occurs for  $\rs\simeq 6.5$ (dash-dotted line), for $\rs=8$ (dotted line) the Mie-frequency of the cluster is already smaller than the laser frequency. $4\times 10^5$ trajectories were shot for uniformly distributed $p_{z,\mathrm{initial}}$ in the range $-3< p_{z,\mathrm{initial}} < 3$.}
  \label{fig:spectra1}
\end{figure}

Photoelectron momentum distributions for different Wigner-Seitz radii $\rs$ are shown in \fref{fig:spectra1}. 
The first observation is that the left ($p_z<0$) and right ($p_z>0$) cut-off momenta change with $\rs$ in a non-monotonous way. They are maximum for $\rs=6.5$, i.e., close to the theoretical value $\rs^{\mathrm{theo}}= 6.75$ for resonance. Hence, not surprisingly, the effect of the RSM-based cluster correction to the SFA is largest at resonance. 
For the unexpanded cluster with $\rs=4$ the spectrum is very similar to the plain-SFA spectrum because the influence of the collective electron motion, modeled by the RSM, is small. For expansions beyond the resonance-$\rs$ the effect of the cluster-correction decreases because the electron sphere excursion decreases.

\begin{figure}
  \includegraphics{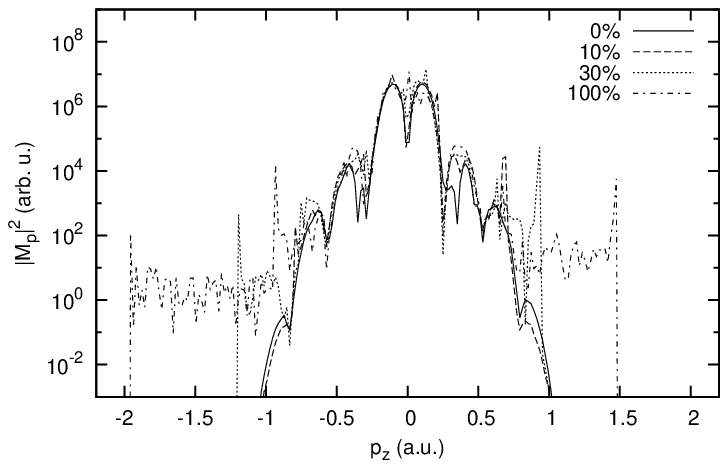}
  \caption{Same as in  \fref{fig:spectra1} for $\rs\simeq 6.5$ but with different prefactors in front of the cluster potential $V_{\mathrm{clu}}(\bi{r},t)$ in \eref{eq:matrixelementtimeintegralS} and \eref{eq:newtonsequationscorrected}. At $0\%$ cluster potential the spectrum is equal to the plain-SFA spectrum. For increasing strength of the potential the cut-off momenta increase, and  a plateau develops.}
  \label{fig:softswitchon}
\end{figure}

\Fref{fig:softswitchon} illustrates how the plateaus develop by increasing the strength $s$ of the  cluster-correction potential,  $V_{\mathrm{clu}} \to s V_{\mathrm{clu}}$, starting from $s=0$ (plain SFA) up to its full value $s=1$. The spikes that appear in the plateaus, especially close to the cut-offs, are  due to the semi-classical nature of the trajectory method and will be discussed below.

\begin{figure}
  \includegraphics{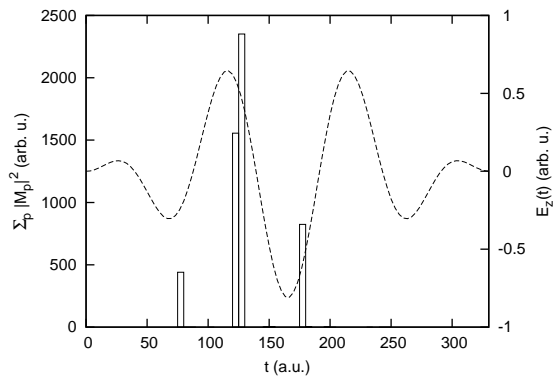}
  \caption{Incoherent sum over $|M_p|^2$  (left axis) for individual trajectories with $|p_{\mathrm{final}}|\geq 0.8$, binned in time. The electric field (dashed) is included (right axis). The majority of all trajectories with significant contributions to this part of the momentum  spectrum is emitted between $t=120$ and $t=130$. These trajectories are selected for further analysis.}
  \label{fig:diagram}
\end{figure}

In order to identify the mechanism by which the fast electrons are generated at resonance  $\rs=6.5$, we first collect the relevant trajectories with $|p_{\mathrm{final}}|\geq 0.8$. \Fref{fig:diagram} shows the incoherent sum of the single-trajectory  probabilities $|M_{\bi{p}}(\ts^{(\alpha)})|^2$ binned in time.  It is seen that the majority of the trajectories contributing to the plateau is emitted within a certain time window around $t=125$. All trajectories starting within this time window have positive final momentum. In fact, the right plateau in  \fref{fig:spectra1} is higher than the left.

\begin{figure}
  \includegraphics{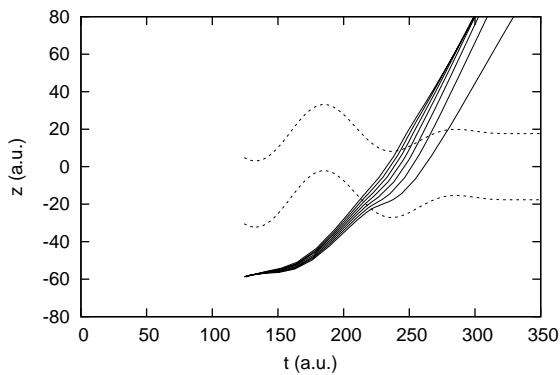}
  \caption{Selected trajectories in position space (every 250th from those emitted between $t=120$ and $130$, see \fref{fig:diagram}). The solid lines are the trajectories $z(t)$, the dotted lines are the boundaries of the oscillating electron cloud.}
  \label{fig:trajectoriesposition}
\end{figure}
\begin{figure}
  \includegraphics{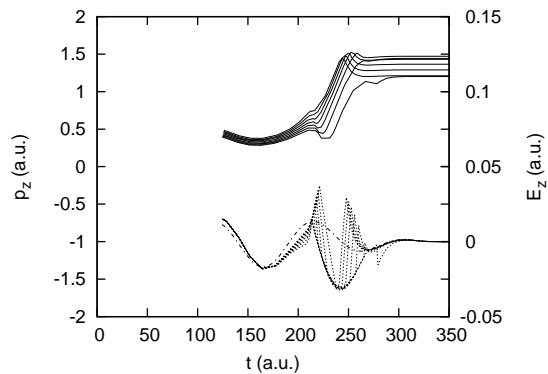}
  \caption{Selected trajectories in momentum space. The solid lines are the canonical momenta $p_z(t)$ (left axis), the dotted lines show the combined electric field $E_z^{\mathrm{total}}(t)$ of laser pulse and cluster potential (right axis) acting on the respective trajectory. The dash-dotted line is the electric field $E_z(t)$ of the laser pulse (right axis). Around $t=230$ the trajectories show a strong acceleration caused by the strongly negative total electric field. Comparing with \fref{fig:trajectoriesposition} this coincides with the additional interaction of the electrons with the cluster.}
  \label{fig:trajectoriesmomentum}
\end{figure}

Trajectories with $p_{\mathrm{final}}\geq 0.8$, emission times $120\leq t_r^{(\alpha)} \leq 130$ and one return to the center of the cluster during the propagation are presented in  \fref{fig:trajectoriesposition} in position space and in \fref{fig:trajectoriesmomentum} in momentum space. \Fref{fig:trajectoriesposition} shows the additional interaction with the cluster. In atomic, plain-SFA slang all these trajectories would be denoted ``long trajectories'': they pass through the origin without rescattering. Within Coulomb-corrected, atomic SFA these trajectories are known to be responsible for the holographic side lobes, for instance \cite{ref_Science_Huismans}. Here, in the cluster-case, the are accelerated by the collective cluster field. This is detailed further in  \fref{fig:trajectoriesmomentum}, where, besides the canonical momentum, the total electric field ``seen'' by the electron is plotted for each of the relevant trajectories.  This field consists of the laser field and the field generated by the cluster potential. While the laser is treated in dipole approximation, the cluster potential is strongly position-dependent. As long as the electron is outside the cluster the total field follows the laser field with only small deviations. Upon entering the cluster  the total field suddenly becomes strongly negative, resulting in significant acceleration of the electrons in positive direction. This is caused by the steep cluster potential in the center at large elongations (see \fref{fig:oscillationpotential}). After leaving the cluster the total field again closely follows the laser field.

Our quantum orbit analysis shows that the acceleration mechanism responsible for the plateau in the spectrum is the additional interaction with the cluster potential in phase with the oscillation of the electron cloud. The magnitude of the acceleration is directly associated with the maximal elongation of the electron cloud, which limits the total field strength during the electron passage. Only electrons arriving at the right phase are accelerated by this process. This leads to the emission of fast electron bunches within narrow time windows, clearly seen in \fref{fig:diagram}. Another observation is that the selected group of trajectories has a peak momentum where the trajectories accumulate, resulting in a sharp peak at the cut-off momentum in the photoelectron spectrum. Such caustics are  known to appear in semi-classical descriptions of quantum dynamics. Although they are typically smoothed in a full quantum treatment, remnants may survive and be observable \cite{ref_PRL_TCSFA}.

\section{Summary}
\label{sec:concl}
We developed a collective-field-corrected strong field approximation (CFSFA) that is capable of explaining energetic electron emission from laser-irradiated metal clusters close to resonance. The approach is basically a combination of the ideas already used for a Coulomb-corrected SFA based on quantum orbits and the rigid-sphere model of cluster physics. The latter is used to estimate the collective field acting on the quantum orbits. So-called ``long trajectories'', well known from the quantum orbit analysis of the plain SFA, that revisit the cluster interior with the right phase are accelerated by the collective field. The advantage of the CFSFA compared to classical calculations is that it treats the ionization step self-consistently. Moreover, the method  allows for interference, which is essential for, e.g.,  the formation of above-threshold ionization peaks, holographic imaging, or any scheme in which structural information is inferred from interference patterns.


\ack
This work was supported by the SFB 652 and project BA 2190/8 of the German Science Foundation (DFG).


\appendix

\section{Force between two overlapping oppositely charged spheres}
\label{app:restoringforce}
Starting from the force on a single charged particle of charge $q$ in a potential $\Phi_2(\bi{r})$ 
\begin{equation}
  \bi{F}(\bi{r})=-q \bnabla \Phi_2(\bi{r}),
\end{equation}
the force on a charge distribution $\rho_1$ (of, e.g., electrons) can be written as
\begin{equation}
  \bi{F}(\bi{r})=-\int \rho_1(\bi{r}')\bnabla_{\bi{r}} \Phi_2(\bi{r}-\bi{r}')\, \rmd^3 r'.
\end{equation}
Taking the divergence gives
\begin{eqnarray}
  \bnabla_{\bi{r}} \cdot \bi{F}(\bi{r}) &=-\bnabla_{\bi{r}} \cdot \left(\int \rho_1(\bi{r}')\bnabla_{\bi{r}} \Phi_2(\bi{r}-\bi{r}')\, \rmd^3 r' \right) \nonumber\\
&= -\int \rho_1(\bi{r}')\Delta_{\bi{r}} \Phi_2(\bi{r}-\bi{r}')\, \rmd^3 r'.
\end{eqnarray}
Using Poisson's equation $\Delta_{\bi{r}} \Phi_2(\bi{r}-\bi{r}')=\Delta_{\bi{r}-\bi{r}'} \Phi_2(\bi{r}-\bi{r}')=-4\pi \rho_2(\bi{r}-\bi{r}')$ where $\rho_2$ is the charge density generating the potential $\Phi_2$, we obtain the overlap integral
\begin{equation}
  \bnabla_{\bi{r}} \cdot \bi{F}(\bi{r})=4\pi \int \rho_1(\bi{r}') \rho_2(\bi{r}-\bi{r}')\, \rmd^3 r'.
\end{equation}
The charge densities for the two spheres in the RSM in atomic units read $\rho_1(\bi{r}')=-n_0 \Theta(R-|\bi{r}'|)$ (electrons) and $\rho_2(\bi{r}-\bi{r}')=n_0 \Theta(R-|\bi{r}-\bi{r}'|)\,$ (ions), with the Heaviside step function $\Theta$ and the particle density $n_0$, which is equal for electrons and ions in the case of a single valence electron per atom.  This reduces the integral to the volume of two spherical caps, which is given by
\begin{equation}
  V = 2\,\frac {\pi h^2}{3} (3R-h)
\end{equation}
where $h=R-r/2$ is the height of each of the two caps. Thus
\begin{eqnarray}
  \nonumber \bnabla_{\bi{r}} \cdot \bi{F}(\bi{r}) &= -4\pi n_0^2 \cdot 2 \frac {\pi (R-\frac{r}{2})^2}{3} \left(3R-(R-\frac{r}{2})\right)\\
  \label{eq:restoringforceradialpartderivative}
  &= -4\pi n_0^2 \frac{4\pi R^3}{3} \left( 1 - \frac{3}{4}\frac{r}{R} +  \frac{1}{16}\frac{r^3}{R^3} \right)\,.
\end{eqnarray}
Whatever direction of $\bi{r}$ is chosen, the dynamics of the two spheres will stay one-dimensional along this direction. For that reason the left hand side can be written as
\begin{eqnarray}
  \bnabla_{\bi{r}} \cdot \bi{F}(\bi{r}) = \frac{1}{r^2}\frac{\partial}{\partial r}[r^2F_r(r)]
\end{eqnarray}
with $F_r$ the radial component of $\bi{F}$. Hence
\begin{eqnarray}
  \label{eq:restoringforcefinalresult}
  F_r(r) &= -4\pi n_0 \frac{4\pi n_0 R^3}{3} \frac{1}{r^2} \int^r \rmd r' r'^2 \left( 1 - \frac{3}{4}\frac{r'}{R} +  \frac{1}{16}\frac{r'^3}{R^3} \right) \nonumber \\
  &= -\frac{4\pi n_0}{3} \frac{4\pi n_0 R^3}{3}   \left( r - \frac{9}{16}\frac{r^2}{R} + \frac{1}{32} \frac{r^4}{R^3} \right) .
\end{eqnarray}
As the ion sphere is much heavier than the electron sphere, we place the ion-sphere center in the origin and introduce the displacement  of the electron-sphere center $\bi{d}$. Taking into account the  vector character of the force $\bi{F}=M\ddot \bi{d}=\frac{4}{3}\pi n_0 R^3\ddot \bi{d}$, $\bi{F}(-\bi{d})=-\bi{F}(\bi{d})$ on the electron cloud of mass $M$ leads to 
\begin{eqnarray}
  \label{eq:displacementeomfinalresult}
  \ddot \bi{d} = -\omegaM^2 \left( \bi{d} - \frac{9}{16R}\bi{d}|\bi{d}| + \frac{1}{32R^3} \bi{d}|\bi{d}|^3 \right), \qquad \omegaM^2=\frac{4\pi n_0}{3},
\end{eqnarray}
which is used in section \ref{sec:rigidspheremodel}, with driver $\bi{E}(t)$ and damping $-\gamma \dot\bi{d}$ added.

\section*{References}

\end{document}